\documentclass[%
 reprint,
 numbers,
 amsmath,amssymb,
 aps,
nofootinbib,
]{revtex4-2}

\usepackage{graphicx}
\usepackage{dcolumn}
\usepackage{bm}
\usepackage{hyperref}
\usepackage{enumitem}
\usepackage{times}
\usepackage{bbm}
\usepackage[normalem]{ulem}
\usepackage{natbib}

\usepackage{xcolor}

\begin{document}


\title{Towards cosmological inference on unlabeled out-of-distribution HI observational data}

\author{Sambatra Andrianomena$^{1,2}$}\email{hagatiana.andrianomena@gmail.com}
\author{Sultan Hassan$^{2,3}$}\email{sultan.hassan@nyu.edu}
\affiliation{%
 $^{1}$SARAO, Liesbeek House, River Park Liesbeek Parkway, Settlers Way, Mowbray, Cape Town, 7705}
 \affiliation{%
 $^{2}$Department of Physics \& Astronomy, University of the Western Cape, Bellville, Cape Town 7535,
South Africa}
  \affiliation{%
 $^{3}$Center for Cosmology and Particle Physics, Department of Physics, New York University, 726 Broadway, New York, NY 10003, USA}




\date{\today}

\begin{abstract}
We present an approach that can be utilized  in order to account for the covariate shift between two datasets of the same observable with different distributions. This helps improve the generalizability of a neural network model trained on in-distribution samples (IDs) when inferring cosmology at the field level on out-of-distribution samples (OODs) of {\it unknown labels}. We make use of HI maps from the two simulation suites in CAMELS, IllustrisTNG and SIMBA. We consider two different techniques, namely adversarial approach and optimal transport, to adapt a target network whose initial weights are those of a source network pre-trained on a labeled dataset. Results show that after adaptation, salient features that are extracted by source and target encoders are well aligned in the embedding space. This indicates that the target encoder has learned the representations of the target domain via the adversarial training and optimal transport. Furthermore, in all scenarios considered in our analyses, the target encoder, which does not have access to any labels ($\Omega_{\rm m}$) during adaptation phase, is able to retrieve the underlying $\Omega_{\rm m}$ from out-of-distribution maps to a great accuracy of $R^{2}$ score $\ge$ 0.9, comparable to the performance of the source encoder trained in a supervised learning setup. We further test the viability of the techniques when only a few out-of-distribution instances are available for training and find that the target encoder still reasonably recovers the matter density. Our approach is critical in extracting information from upcoming large scale surveys.
\end{abstract}

\maketitle



\section{\label{sec:introduction}INTRODUCTION}
Previous studies showed that HI intensity mapping technique \citep{battye2004neutral, chang2008baryon}, which consists of collecting the combined emission of HI galaxies of large scale structure at 21cm wavelength, can be used as a powerful probe to constrain cosmological parameters \citep[e.g.][]{bull2015late, pourtsidou2017h}. Several ongoing and upcoming HI experiments are expected to provide more insights into the late time acceleration of the Universe, such as Square Kilometre Array \citep[SKA; ][]{mellema2013reionization}, BINGO \citep{battye2013h}, the Canadian Hydrogen Intensity Mapping Experiment \citep[CHIME; ][]{bandura2014canadian}, Five-hundred-meter Aperture Spherical radio Telescope \citep[FAST; ][]{bigot2015hi, li2023fast}, Hydrogen Epoch of Reionization Array \citep[HERA; ][]{deboer2017hydrogen}, MeerKAT Large Area Synoptic Survey \citep[MeerKLASS; ][]{santos2017meerklass}.  

By utilizing 21 cm power spectrum in a Bayesian framework (such as Monte Carlo Markov Chain), various studies showed that constraining cosmological parameters to a promising accuracy is achievable  \citep[e.g.][]{greig201521cmmc, hassan2017epoch}. Nevertheless, power spectrum analysis are prone to information loss since it mainly encodes the mean distribution of matter clustering at different $k$ modes. Previous works for instance had to resort to high-order statistics (such as bispectrum) \citep{bharadwaj2005probing, majumdar2018quantifying} to further capture the non-gaussianity in 21 cm signal. This helps break the degeneracies between parameters, improving the constraints obtained from HI intensity mapping surveys \citep{randrianjanahary2024cosmological}. In order to circumvent the challenges that come with modeling physics at non-linear scales, and to minimize the signal loss inherent in power spectrum analyses, extracting the parameters at the field level (i.e. two dimensional maps or three dimensional data cube) using deep learning looks very promising. For instance, \cite{gillet2019deep} trained a convolutional neural network (CNN) to predict astrophysical parameters from 2D HI maps to a great accuracy; \cite{hassan2022hiflow} made use of HI maps from CAMELS dataset \citep{villaescusa2021camels} to train a Masked Autoregressive Flow \citep[MAF;][]{papamakarios2017masked} which successfully learned the HI data distribution so as to able to generate new images mimicking the training data. By conditioning on the matter density ($\Omega_{\rm m}$) and the amplitude of the density fluctuations ($\sigma_{8}$), and using the marginal likelihood which is obtained by inverting the flow, the model in \cite{hassan2022hiflow} can infer the underlying cosmology of HI maps.
\cite{andrianomena2023predictive} also demonstrated that the constraints on the cosmology and astrophysics improved by combining HI maps with the gas density and amplitude of the magnetic fields maps at the input of their CNN model. The latter also output the predictive uncertainties on each inferred parameter by using Bayesian approximation. \cite{hassan2020constraining} highlighted the fact that the performance of their deep networks degraded when trained to recover cosmology and astrophysics from noisy (more realistic) HI maps. 

Foreground noise, whose dominant component is the galactic synchrotron emission, buries the cosmological signal, This makes information extraction challenging. \cite{liu2009improved}, for instance, modeled the noise contamination in Fourier space before removing it. \cite{liu2011method} prescribed a unified matrix-based approach to subtract the foregrounds and estimate the power spectrum of the signal. Provided the smoothness in frequency of the contaminating noise, \cite{alonso2015blind} opted for a blind signal separation technique which consists of removing the most dominant component of the foregrounds by using Principal Component Analysis \citep[PCA;][]{wold1987principal}. Deep learning based methods have also been explored in order to clean the noisy maps. In their analyses, \cite{makinen2021deep21} built a denoising U-net model to predict the cosmological signal from PCA-reduced maps, i.e. maps whose dominant contributions were removed using PCA.

Despite various noise prescriptions adopted when simulating HI maps for a given survey, the mock datasets for training  neural network models will always be different from the real datasets collected from the experiment. This data shift, i.e. difference between the distributions of real and simulated data, is most likely to hit the performance of a deep learning model trained on the simulated maps when tested on dataset from surveys. In order to overcome this issue, domain adaptation (DA) techniques, which mainly consist of learning robust representations that are invariant to domain shift, have been proposed \citep[e.g.][]{ganin2015unsupervised, sun2016deep, bousmalis2016domain}.  
Various DA approaches have been explored to address issues in Astronomy and Astrophysics. \cite{dell2023deep} considered three different deep unsupervised domain adaptation methods to improve the robustness of their deep network in preparation for Imaging Atmospheric Cherenkov Telescopes event reconstruction. Semi-supervised universal DA was used to account for the varying systematics in different galaxy surveys within the context of galaxy morphology classification and anomaly detection \citep{ciprijanovic2023deepastrouda} \citep[for the use of DA galaxy morphology classification, see also][]{ciprijanovic2022deepadversaries}. 
\cite{gilda2024unsupervised} opted for Kullback–Leibler Importance Estimation Procedure \citep{sugiyama2007direct} to constrain star formation history. \cite{ciprijanovic2020domain} used Maximum Mean Discrepancy \citep[MMD;][]{smola2006maximum} and Domain Adversarial Neural Networks \citep[DANN;][]{ganin2015unsupervised} to improve the performance of their network in classifying galaxy mergers.

This far, several deep networks have been successfully trained on HI mock data to recover the underlying cosmology. However there are two issues that need to be addressed in a real-world scenario
\begin{itemize}
    \item the complexity of noise modeling in HI experiments leads to a domain shift which will negatively impact the performance of a model trained on simulated data,
    \item realistic scenario, i.e. HI maps from experiments, will not have labels that can be used to evaluate the model predictions. 
\end{itemize}
In this proof of concept, we propose an approach that can potentially tackle those problems by making use of unsupervised domain adaptation techniques. The latter will help improve the generalizability of a deep network model when attempting to extract cosmological information from real data of HI surveys. We present the datasets considered in our analyses in \S\ref{sec:data-simulations}, and describe the two approaches in \S\ref{sec:method}. We show the results in \S\ref{sec:results} and conclude in \S\ref{sec:conclusion}.


\section{Data}\label{sec:data-simulations}
In this work we consider the multifield dataset \citep{villaescusa2022camels} from the CAMELS Project \citep{villaescusa2021camels, villaescusa2022camels1}. CAMELS uses two different state-of-the-art hydrodynamic simulation suites, IllustrisTNG \citep{nelson2019illustristng} and SIMBA \citep{dave2019simba}, to generate hundreds of thousands of 2D field maps which correspond to a region of $25\times25~(h^{-1}{\rm Mpc})^2$ at $z=0$. The maps, with a resolution of $256\times256$ pixels, were obtained by varying the cosmology ($\Omega_{m}$, $\sigma_{8}$) and the astrophysics composed of the stellar feedbacks ($A_{\rm SN1}$, $A_{\rm SN2}$) and active galactic nuclei (AGN) feedbacks ($A_{\rm AGN1}$, $A_{\rm AGN2}$).

\begin{figure}
\includegraphics[width=0.48\textwidth]{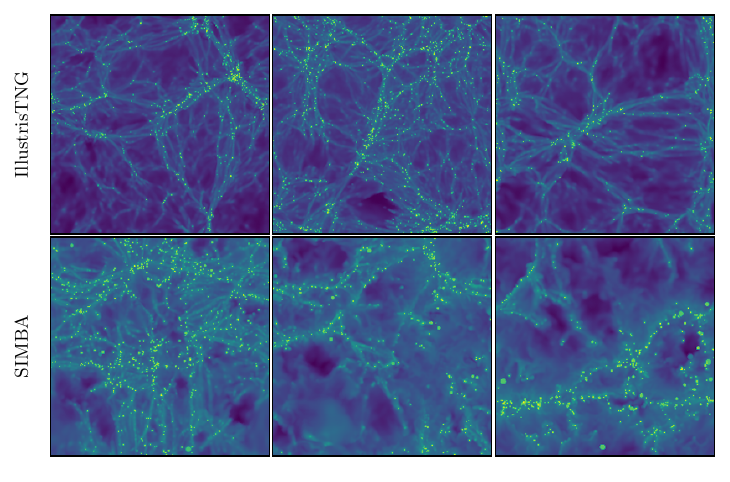}
\caption{\label{fig:images} Random samples of HI maps from IllustrisTNG and SIMBA are shown in the top and bottom panels respectively. The color coding in both datasets is set to be the same for a better comparison. This figure shows that the HI maps are visually very different in the two simulations.}
\end{figure}

\begin{figure}
\includegraphics[width=0.45\textwidth]{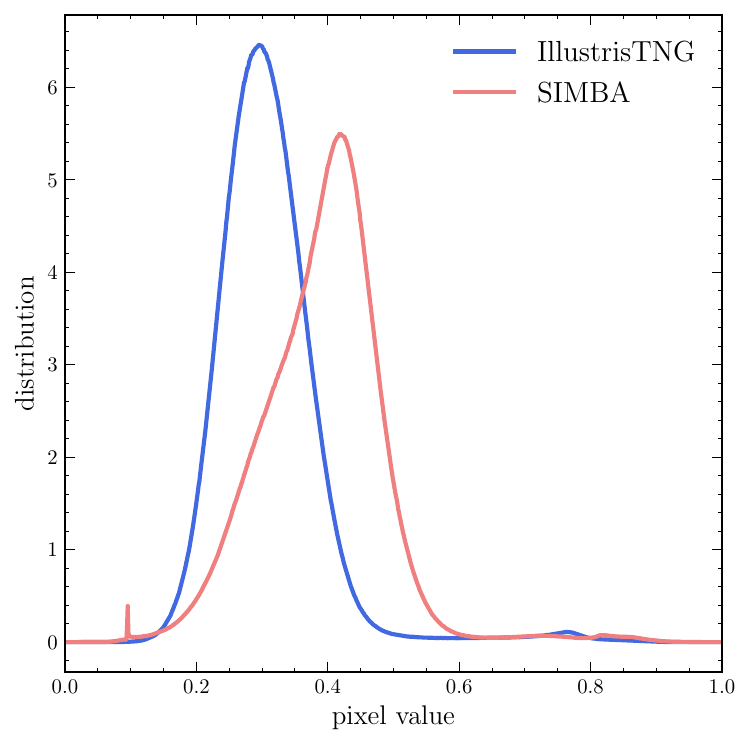}
\caption{\label{fig:pdf} Probability distribution of the pixel values of the HI maps from each simulation. Blue and red solid lines denote the distributions from IllustrisTNG and SIMBA respectively. The pixel values in all maps are normalized to [0,1]. This figure shows a clear distribution shift between SIMBA and IllustrisTNG at the level of HI maps.}
\end{figure}

To mimic the source and target domains in our analyses, we make use of the HI maps from SIMBA and IllustrisTNG which are two datasets generated by two different distributions owing to the different ionizing background adopted in each suite (SIMBA uses the treatment prescribed in \cite{haardt2012radiative} whereas IllustrisTNG opts for the method described in \cite{faucher2009new}). We present in Figure~\ref{fig:images} samples from both datasets. All the topological features, e.g. voids, filaments and halos, are noticeable in both types of samples, but the clear difference lies in the image texture. The small scale features in SIMBA (see bottom panels in Figure~\ref{fig:images}) tend to be a bit blurry in comparison with those of IllustrisTNG (see top panels in Figure~\ref{fig:images}). To further highlight the difference between the two datasets, we plot in Figure~\ref{fig:pdf} the distributions of the mean value of the pixels of the two domain images. After normalizing the pixel values in all maps to [0,1], each curve is obtained by averaging the pixel value distribution of all the images in each domain. Red and blue solid lines denote SIMBA and IllustrisTNG respectively. There is a clear shift between the two distributions in Figure \ref{fig:pdf}, suggesting that the samples in one dataset are viewed as out-of-distribution samples of the other. We first treat SIMBA and IllustrisTNG as the source and target domains respectively, then vice versa. In what follows, unless otherwise stated, we will use source  and target domain maps/data interchangeably with in-distribution and out-of-distribution maps/data respectively.

In this regression task, the networks are trained to extract the underlying cosmology ($\Omega_{\rm m}$ and $\sigma_{8}$) of the maps, while marginalizing over the astrophysical parameters. The prior ranges of the labels, i.e. the cosmological parameters, are the same in both domains, the matter density $\Omega_{\rm m}$ within [0.1, 0.5] and the amplitude of the density fluctuations $\sigma_{8}$ within [0.6, 1.0] \citep{villaescusa2022camels}. We select 12,000 pairs of \{map, label\} from each domain to train the two networks, and 1,500 of unseen instances from both domains are used for testing. It is worth noting that the labels of the target domain are not accessed during training phases, i.e. pre-training of the source network and adaptation of the target network, but are only used to assess the performance of the adapted target encoder. 

\section{Method}\label{sec:method}
We consider two methods that have similar training steps and use the exact same source and target encoders, but are different in how the target encoder weights are updated during the adaptation phase. In the following subsections, we first provide a very brief introduction of domain adaptation, especially the case relevant to our study, and then present the two methods.
\subsection{\label{subsec:domain-adaptation}Domain adaptation}
A domain $\mathcal{D}$ is composed of a feature space $\mathcal{X}$ whose elements $x$ are in $d$-dimensions (i.e. $x \in \mathbb{R}^{d}$), a label space $\mathcal{Y}$ whose elements can be either a class label in the case of classification or a dependent variable (scalar) defined with a range (i.e. prior) in the case of regression, and finally a join probability distribution $p(x,y)$. Since, in a supervised learning, the label is inferred from the input feature, the join probability distribution can be written as $p(x,y) = p(y|x)p(x)$\footnote{It can also be written as $p(x,y) = p(x|y)p(y)$ but we choose the one in the text to be consistent with a standard supervised learning setup where the aim is to predict the label.} where the first term is the  probability of the label given the input and the second term is known as the prior. 

In general for a given supervised learning task, the test set\footnote{Or generally the target domain.} is assumed to be from the same distribution of the dataset\footnote{Or also the source domain.} that is used to train a model. However, in some cases during deployment of the trained model, its performance significantly degrades as the assumption is no longer valid, i.e. there is a shift between the source and target domains. The aim of domain adaptation is to minimize the generalization error of a trained model in the presence of a domain shift which can be \citep[for a review, see][]{farahani2021brief}
\begin{itemize}
    \item a prior shift which arises when the label distribution of the target domain differs from that of the source domain, in other words $p_{S}(y_{S})\ne p_{T}(y_{T})$,
    \item covariate shift characterized by a discrepancy between the feature space of source domain and that of the target domain, i.e. $p_{S}(x_{S})\ne p_{T}(x_{T})$, 
    \item concept shift where the conditional distributions in source and target domains are different, i.e. $p_{S}(y_{S}|x_{S})\ne p_{T}(y_{T}|x_{T})$. 
\end{itemize}
In our analyses, it is assumed that \textit{the labels of the target domain do not exist}, a case of an unsupervised domain adaptation. This is to mimic a real-world scenario where we infer the cosmological parameters of maps from experiments using a deep learning model trained on simulated data. We also assume that the label space and the mapping between feature and label remain the same in both domains, so that we deal with a covariate shift in our investigation. These assumptions are valid since the same range of cosmological parameters (i.e. prior range) is adopted in both simulation suites ($p_{S}(y) = p_{T}(y)$), and regardless of the treatment of the baryon physics used in SIMBA and IllustrisTNG, the topological features of the maps in both cases are equally sensitive to the cosmology. We consider two approaches in our study which will be presented in the following sections.


\begin{figure*}
\includegraphics[width=0.8\textwidth]{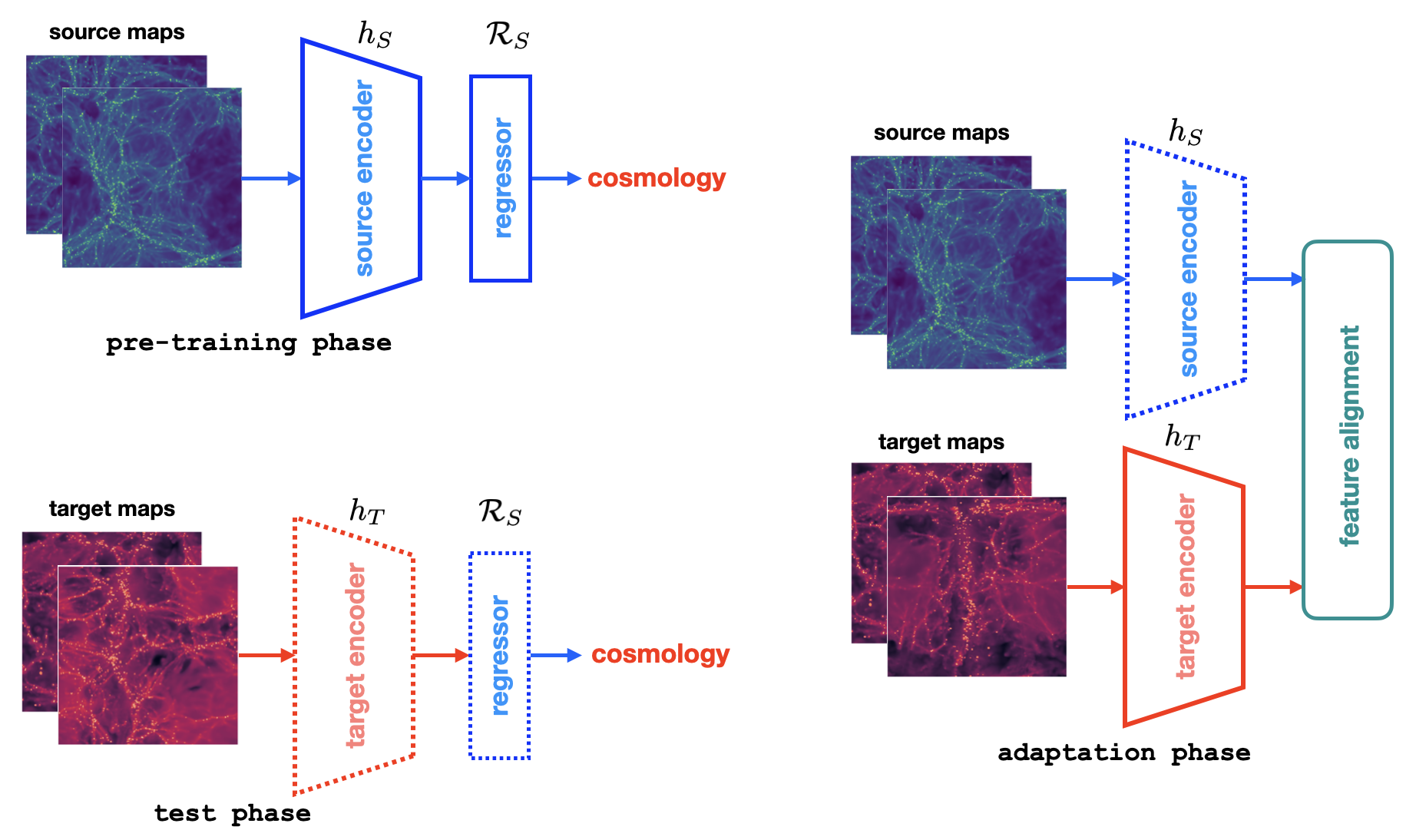}
\caption{\label{fig:adversarial-method} Schematic view of the different phases. The source encoder and a regressor are first trained to infer underlying cosmology of in-distribution instances. At the beginning of the adaptation, the weights of the pre-trained source encoder are used to initialize the target encoder (whose architecture is identical to that of the source feature extractor). The target encoder is trained by aligning the features it extracts from the target maps with those of the source maps learned by the source encoder. Finally, during testing, the representations of the test set from the target domain, which are encoded by the trained target encoder, are fed to the regressor for inference. It is noted that the weights of a component in dashed line (blue or red) are frozen during a given phase. }
\end{figure*}
\subsection{\label{subsec:adda} Adversarial discriminative domain adaptation} 
We first consider an adversarial method, named Adversarial Discriminative Domain Adaptation (ADDA), which was prescribed in \cite{tzeng2017adversarial} for a classification task originally. The setup consists of feature $\mathcal{X}_{S}$ and label $\mathcal{Y}_{S}$ spaces of the source domain and feature $\mathcal{X}_{T}$ of the target domain which is unlabeled. In other words, we do not have access to the label space $\mathcal{Y}_{T}$ during the training phases. The goal is to be able to predict the labels corresponding to the inputs from the target domain maps. 

The idea is to first learn representations $\bm{h}_{S}(\mathcal{X}_{S})$ of the in-distribution images along with a regressor $\mathcal{R}_{S}$ that is used to recover the underlying cosmology ($\Omega_{\rm m},\>\sigma_{8}$) of the inputs (maps) in a supervised learning setup. The source encoder $\bm{h}_{S}$, shown in Figure~\ref{fig:adversarial-method} (pre-training phase), has an architecture similar to a VGG feature extractor \citep{simonyan2014very} composed of different stages that are built from chaining up 2 or 3 convolutional layers (see Table~\ref{encoder_arch} for more details). The outputs of the last layer in each stage is halved using max pooling layers (with stride = 2) and from one stage to another the number of channels is doubled. A single fully connected layer of 512 units constitutes a regressor $\mathcal{R}_{S}$. The loss that is considered in the pre-training phase is the same as the one used in \cite{andrianomena2023predictive} 
\begin{eqnarray}\label{eq:loss-function}
    &&\mathcal{L}_{Reg} = \frac{1}{N\times M}\sum_{i = 1}^{N}\Bigg[{\rm log}\left(\sum_{j = 1}^{M}(y_{i,j} - \omega_{i,j})^{2}\right) \nonumber\\
   &&~~~~ + {\rm log}\left(\sum_{j = 1}^{M}\left((y_{i,j} - \omega_{i,j})^{2} - \sigma_{i,j}^{2}\right)^{2}\right)\Bigg],
\end{eqnarray}
where $N$, $M$, $y_{i,j}$, $\omega_{i,j}$ and $\sigma_{i,j}$ are the number of predicted parameters ($\Omega_{\rm m},\sigma_{8}$), batch size, ground truth, prediction and standard deviation respectively. The cosmological parameters ($\Omega_{\rm m},\sigma_{8}$) and their corresponding standard deviation are the outputs of the regressor\footnote{This implies that the total number of outputs of the regressor is 4.}.  

The adaptation phase (Figure~\ref{fig:adversarial-method}) consists of training adversarially a target encoder $\bm{h}_{T}$, whose architecture is identical to that of $\bm{h}_{S}$, such that the representations of target domain maps $\bm{h}_{T}(\mathcal{X}_{T})$ are well aligned with $\bm{h}_{S}(\mathcal{X}_{S})$ in a lower dimensional latent space. Initially, the weights of the pre-trained source encoder are transferred to the target encoder, in order to avoid divergence during training. During adaptation, the source encoder weights are frozen, and a discriminator $\mathcal{C}$ (see Table~\ref{discriminator_arch} for more details),  which comprises two hidden fully connected layers of 512 units each, learns to separate the two latent codes $\bm{h}_{S}(\mathcal{X}_{S})$ and $\bm{h}_{T}(\mathcal{X}_{T})$ (see Figure~\ref{fig:adversarial-method}). In order to align the two different representations, the loss is composed of \citep{tzeng2017adversarial}
\begin{eqnarray}\label{eq:adv-C}
    &&\underset{D}{\rm min}\>\mathcal{L}^{\mathcal{C}}_{{\rm adv}}  = - \mathbb{E}_{x_{S}\sim p_{S}(x_{S})}[{\rm log}\>\mathcal{C}(\bm{h}_{S}(\mathcal{X}_{S}))]\nonumber \\
    &&~~~~-\mathbb{E}_{x_{T}\sim p_{T}(x_{T})}[{\rm log}(1 - \mathcal{C}(\bm{h}_{T}(\mathcal{X}_{T})))],
\end{eqnarray}
which is a component that is used to update the discriminator such that its ability to differentiate between the two representations improves during training, and \citep{tzeng2017adversarial}
\begin{equation}\label{eq:adv-hS}
\underset{\bm{h}_{S},\bm{h}_{T}}{\rm min}\>\mathcal{L}^{\bm{h}_{T}}_{{\rm adv}}  = -\mathbb{E}_{x_{T}\sim p_{T}(x_{T})}[{\rm log}\>\mathcal{C}(\bm{h}_{T}(\mathcal{X}_{T}))]
\end{equation}
which is used to update to target encoder. Another form of the adversarial loss in Equation~\ref{eq:adv-hS} can be obtained by simply taking the negative of the standard loss in Equation~\ref{eq:adv-C}, i.e. $\mathcal{L}^{\bm{h}_{T}}_{{\rm adv}} = -\mathcal{L}^{\mathcal{C}}_{{\rm adv}}$. Practically, this approach requires a gradient reversal layer \citep[GRL;][]{ganin2015unsupervised} which acts as an identity function during forward pass but multiplies the gradient by -1 during backward pass, amounting to a gradient ascent (maximization). The GRL can help the discriminator training converge quickly but however is more likely to cause vanishing gradient \citep{tzeng2017adversarial}. 

It is worth pointing out that the target feature extractor plays a role similar to that of a generator in GANs by trying to fool the discriminator. To retrieve the cosmological information from a test set of the target domain, the pre-trained regressor $\mathcal{R}_{S}$\footnote{The one used for prediction in pre-training phase.} is fed with the representations extracted by the adapted network (see Figure~\ref{fig:adversarial-method} test phase).
\begin{table*}
\caption{$R^{2}$ score achieved on $\Omega_{\rm m}$ and $\sigma_{8}$ from predicting the source and target datasets using the source and trained target network respectively. SN and TN denote source and target networks respectively. IDs and OODs in the table refer to in-distribution (source) and out-of-distribution (target) samples. Parameter inference improves significantly after adaptation (see SN (OODs) vs TN (OODs) columns).}
  \centering
  \begin{tabular}{| *{9}{c|} }
    \hline
    & \multicolumn{2}{c|}{SN (IDs)} & \multicolumn{2}{c|}{SN (OODs)} & \multicolumn{2}{c|}{ $\rm TN^{ADDA}$ (OODs)} & \multicolumn{2}{c|}{$\rm TN^{OT}$ (OODs)} \\\cline{2-9}
    
    & $\Omega_{\rm m}$ & $\sigma_{8}$ & $\Omega_{\rm m}$ & $\sigma_{8}$ & $\Omega_{\rm m}$ & $\sigma_{8}$ &  $\Omega_{\rm m}$ & $\sigma_{8}$   \\
    \hline
    SIMBA$\rightarrow$TNG & 0.948 & 0.720 & -0.474 & -2.481 & 0.945 & 0.735 &  0.947 & 0.394   \\
    \hline
    TNG$\rightarrow$SIMBA & 0.974 & 0.884 & -2.109 & -2.112 & 0.903 & 0.205 & 0.924 & -0.215   \\
    \hline
  \end{tabular}
\label{tab:rsquare}
\end{table*}
\begin{figure}
\includegraphics[width=0.48\textwidth]{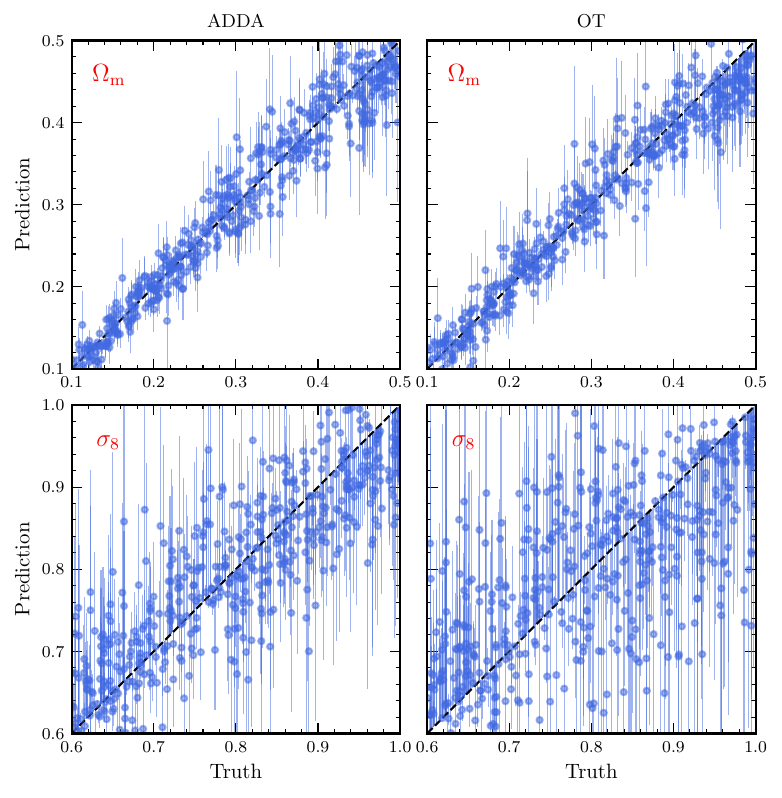}
\caption{\label{fig:sim-tng-prediction} For the SIMBA$\rightarrow$TNG experiment, predictions on the out-of-distribution test set are denoted by blue dots, whereas the error bars indicate the absolute difference between the ground truth and predicted parameter. Results shown in the first and second columns are obtained from the target network adapted using ADDA and OT methods respectively.}
\end{figure}
\begin{figure*}
\includegraphics[width=0.72\textwidth]{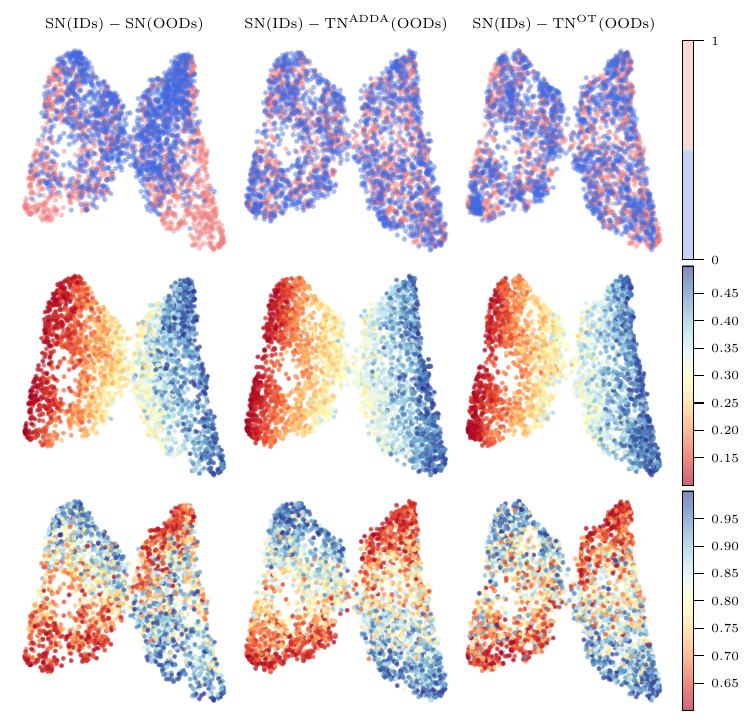}
\caption{\label{fig:sim-tng} The source and target datasets are SIMBA and IllustrisTNG respectively. 2D visualization of the features extracted by the source and target encoders using UMAP. \textit{Top row}: in all panels the red dots denote the representations of in-distribution maps. The blue dots in first, second and third panels indicate the representations of the target dataset learned by source encoder, trained target encoder using ADDA and trained target encoder using optimal transport respectively. It can be clearly noticed that the representations are well aligned after adaptation. \textit{Middle row}: each panel in upper row is plotted again but using the corresponding value of $\Omega_{\rm m}$ for the color coding. \textit{Bottom row}: the color coding is related to the value of $\sigma_{8}$}.
\end{figure*}
\begin{figure}
\includegraphics[width=0.48\textwidth]{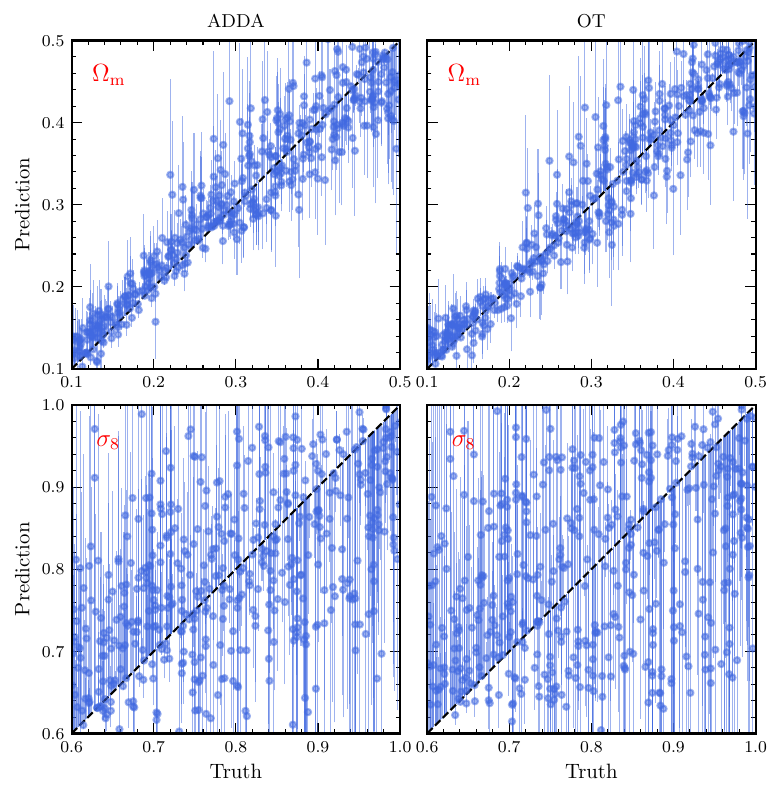}
\caption{\label{fig:tng-sim-prediction} Similar to what is shown in Figure~\ref{fig:sim-tng-prediction} but for TNG$\rightarrow$SIMBA setup.}
\end{figure}
\begin{figure*}
\includegraphics[width=0.72\textwidth]{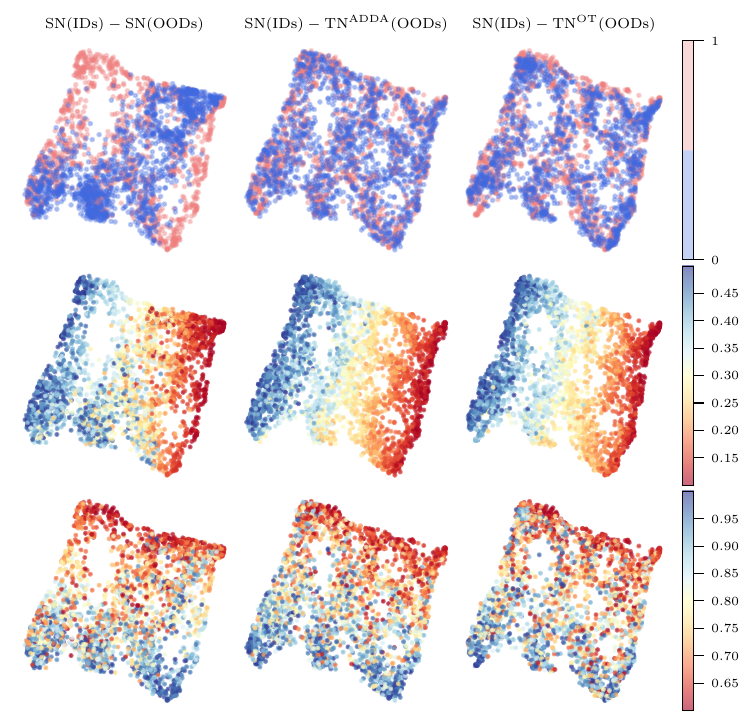}
\caption{\label{fig:tng-sim} Similar to what is shown in Figure~\ref{fig:sim-tng} but the source and target datasets are IllustrisTNG and SIMBA respectively. The top panels show that there is a good overlap between the two representations after adaptation.}
\end{figure*}
We pre-train both the source feature extractor and regressor for 200 epochs, using Adam optimizer with learning = 0.0015 and opting for a batch size = 50. The learning rate is halved whenever the loss function does not improve over 5 epochs by using \texttt{ReduceOnPlateau}. The hyperparameters for the pre-training are based on the prescription in \cite{andrianomena2023predictive}. The adaptation phase is run for 200 epochs in the case where IllustrisTNG constitutes the out-of-distribution dataset, and 800 epochs when SIMBA is target domain. The reason for choosing different epochs for the two scenarios will be more apparent in \S\ref{sec:results}. The target encoder and the classifier have their own Adam optimizer with learning rates $10^{-5}$ and $10^{-4}$ respectively. The values of the learning rates are from this \href{https://github.com/corenel/pytorch-adda/tree/master}{implementation}\footnote{An implementation of the method prescribed in \cite{tzeng2017adversarial} can be found in \href{https://github.com/corenel/pytorch-adda/tree/master}{https://github.com/corenel/pytorch-adda/tree/master}} upon which ours\footnote{\href{https://github.com/hagasam/domain-adaptation-cosmo/tree/main}{https://github.com/hagasam/domain-adaptation-cosmo/tree/main}} is based. It is worth reiterating that the pre-trained regressor is only used when testing the performance of the adapted target encoder.

\subsection{\label{sec:optimal-transport} Optimal transport}
Optimal Transport \citep[OT;][]{courty2016optimal} is an approach whose objective is to find an optimized solution to move unit mass between different distributions. It can also be used to estimate the dissimilarity between two probability distributions, similar to Maximum Mean Discrepancy \citep{smola2006maximum}. Generally, within the context of domain adaptation problems, OT approach consists of finding a transport map $\mathcal{T}$, which transforms samples $x$ from the source domain to the target domain, such that the transportation cost $\bm{C}$ \citep{courty2016optimal}
\begin{equation}
    \bm{C}(\mathcal{T}) = \int c(x, \mathcal{T}(x))d\mu(x),
\end{equation}
is minimized, according to \citep{courty2016optimal} 
\begin{equation}
    \mathcal{T}_{0} = \underset{\mathcal{T}}{\rm argmin}\int c(x, \mathcal{T}(x))d\mu(x),
\end{equation}
where $c(.,.)$ is a cost function and $\mu$ is a marginal distribution. This optimal transport problem can be framed differently as \citep{kantorovitch1958translocation}
\begin{equation}
    \gamma_{0} = \underset{\gamma \in \mathbb{R}_{+}^{n_{S}\times n_{T}}}{\rm argmin}\int \int c(x_{S},x_{T})\gamma(x_{S},x_{T}){\rm d}x_{S}{\rm d}x_{T},
\end{equation}
where $\gamma$, known as the transport plan, is a joint probability and denotes the amount of mass to be moved between the two distributions. $n_{S}$ and $n_{T}$ are the numbers of samples in the source and target domains respectively. In the discrete case, we have that
\begin{equation}
    \gamma_{0} = \underset{\gamma \in \mathcal{B}}{\rm argmin} \underset{i,j}{\sum}\gamma_{i,j}{\rm M}_{i,j},
\end{equation}
such that \citep{courty2016optimal}
\begin{equation}
    \mathcal{B} = \{\gamma \in \mathbb{R}_{+}^{n_{S}\times n_{T}}\> |\> \gamma\bm{1_{\rm n_{S}}}=\mu_{s}, \gamma\bm{1^{\rm T}_{\rm n_{t}}}=\mu_{t}\},
\end{equation}
and ${\rm M}_{i,j}$ is a metric cost matrix. To align the two different distributions of representations $\mu_{s}$, $\mu_{t}$ of source and target domains respectively, we minimize the Wasserstein distance \citep{bonneel2011displacement}, also known as Earth Mover's Distance \citep{rubner2000earth}, given by
\begin{equation}\label{emd2}
    W_{p}(\mu_{s},\mu_{t}) = \Bigg(\underset{\gamma \in \mathcal{B}}{\rm argmin} \underset{i,j}{\sum}\gamma_{i,j}||\bm{h}_{S}(x_i)-\bm{h}_{T}(x_j)||_{p}\Bigg)^{\frac{1}{p}},
\end{equation}
where we opt for $p = 2$ and $x_{i}$, $x_{j}$ are samples from $\mathcal{X}_{S}$ and $\mathcal{X}_{T}$ respectively. In other words, Equation~\ref{emd2} is the loss function for feature alignment we consider when using OT method as opposed to the adversarial losses in Equations~\ref{eq:adv-C} and \ref{eq:adv-hS}.

During adaptation phase using OT method, the source and target encoders are respectively fed with source and target domain instances. The OT loss is computed by using the latent codes from the two feature exctractors. Finally, only the target encoder is updated by using the OT loss at each step of the training. The adaptation is run for 200 epochs with a learning rate $10^{-5}$. We utilize the library developed in \cite{flamary2021pot} in our implementation of the loss function. The pre-trained regressor is only used to assess the performance of the adapted target encoder during testing.

It is worth noting that in a given setup e.g. SIMBA and IllustrisTNG are in- and out-of-distribution datasets respectively, the pre-training phase is only run once. Each feature alignment method (ADDA and OT) is run separately using the optimized weights from the pre-trained source encoder to initialize the target encoder used in each method. Similarly, the same pre-trained regressor is used to assess the performance of the target encoders adapted from the two methods. 


\section{Results}\label{sec:results}
\subsection{SIMBA (IDs)$\rightarrow$IllustrisTNG (OODs)}\label{subsec:sim-tng}
We first choose SIMBA HI and IllustrisTNG HI datasets to be the source and target domains respectively (SIMBA $\rightarrow$ TNG). It is worth reiterating that during the two training phases, pre-training and adaptation, the labels ($\Omega_{\rm m},\sigma_{8}$) associated with the out-of-distribution maps (IllustrisTNG) are not used at all. The latter are only used to validate the predictions of the adapted encoder $\bm{h}_{T}$.

To evaluate the accuracy of our model predictions, we select the coefficient of determination $R^{2}$ which  explains both the variance of the predicted parameters and the strength of correlation between the ground truth and prediction. We have that  
\begin{equation}\label{r2score}
    R^{2} = 1 - \frac{\sum_{i = 1}^{n}\left(y_{i}-\omega_{i}\right)^{2}}{\sum_{i = 1}^{n}\left(y_{i}-\bar{y}\right)^{2}},
\end{equation}
where $n$ and $\bar{y}$ are the number of samples in the test set and mean of the ground truth respectively. We
\begin{table}
\caption{$R^{2}$ score achieved on $\Omega_{\rm m}$ and $\sigma_{8}$ similar to Table~\ref{tab:rsquare} but results are obtained from adapting the target encoder by using only 100 instances from the target domain.}
  \centering
  \begin{tabular}{| *{5}{c|} }
    \hline
    & \multicolumn{2}{c|}{ $\rm TN^{ADDA}$(OODs)} & \multicolumn{2}{c|}{$\rm TN^{OT}$(OODs)} \\\cline{2-5}
    & $\Omega_{\rm m}$ & $\sigma_{8}$ &  $\Omega_{\rm m}$ & $\sigma_{8}$   \\
    \hline
    SIMBA$\rightarrow$TNG & 0.891 & 0.616 &  0.895 & 0.611   \\
    \hline
    TNG$\rightarrow$SIMBA & 0.698 & 0.111 & 0.692 & 0.299   \\
    \hline
  \end{tabular}
\label{tab:rsquare-small-samples}
\end{table}
present in Table~\ref{tab:rsquare} first row the performance of the model for the scenario SIMBA$\rightarrow$TNG. Results show that the source network\footnote{It is worth mentioning that in the text "network" refers to "encoder + regressor".} (i.e. $\bm{h}_{S}+\mathcal{R}_{S}$) is able to constrain the matter density to a good accuracy with $R^{2}$ = 0.948 on in-distribution maps. Although predicting $\sigma_{8}$ seems more challenging, the deep regressor still achieves $R^{2}$ = 0.72. This can be accounted for by the fact that the HI maps are more sensitive to matter density as neutral hydrogen is used as a bias tracer of dark matter. They are related by a bias relation 
$\delta_{\rm HI} \sim b_{\rm HI}\delta_{\rm m}$, where $\delta_{\rm HI}$, $\delta_{\rm m}$ and $b_{\rm HI}$ are HI density, matter density and bias factor respectively \citep{wang2021breakdown}. We make predictions on the out-of-distribution images (IllustrisTNG) using the pre-trained source network to check how much the data shift impacts its performance and find that it fails to recover the cosmology, with $R^{2}$ = -0.474, -2.481 for $\Omega_{\rm m}$ and $\sigma_{8}$ respectively. After adaptation, the target encoder is used to predict the cosmology of HI IllustrisTNG maps. Using ADDA method for the feature alignment, the target network TN$^{\rm ADDA}$ (i.e. $\bm{h}_{T}$ + $\mathcal{R}_{S}$) achieves very promising performance on the test set from the target domain -- $R^{2}$ = 0.945, 0.735 on $\Omega_{\rm m}$ and $\sigma_{8}$ respectively -- similar to that of the source network SN on HI SIMBA maps (see Table~\ref{tab:rsquare} first row). This demonstrates the robustness of the adaptation technique which does not access the labels of the out-of-distribution dataset at all during the training and uses the regressor $\mathcal{R}_{S}$ pre-trained on the source domain for inference. On the other hand, the adapted network TN$^{\rm OT}$, whose feature alignment is achieved using optimal transport, constrains the matter density on the target domain to a precision comparable to that of the source network on the source domain, but fails to extract the amplitude of the density fluctuations, $R^{2}$ = 0.394 on $\sigma_{8}$. This indicates that in this specific scenario, SIMBA$\rightarrow$TNG, ADDA method is more robust than the optimal transport. We also present in Figure~\ref{fig:sim-tng-prediction} how well the target network, adapted by either ADDA or OT methods, is able to constrain the cosmological parameters. The blue dots denote the network predictions and the corresponding error bars are related to the absolute difference between the ground truth and predicted parameters. Regardless of the method used for feature alignment, it can be seen from Figure~\ref{fig:sim-tng-prediction} that most of the $\Omega_{\rm m}$ predictions are relatively close the identity line (dashed black line). This suggests that matter density can be well recovered from the out-of-distribution HI maps, consistent with the $R^{2}$ values for $\Omega_{\rm m}$ obtained in Table~\ref{tab:rsquare}. However, the $\sigma_{8}$ predictions (see bottom panels in Figure~\ref{fig:sim-tng-prediction}) are further from the identity line, which translates to larger error bars. By comparing the two bottom panels in Figure~\ref{fig:sim-tng-prediction}, it is noticeable that the scatter (about the identity line) in TN$^{\rm OT}$ predictions (right panel) is larger. In other words, TN$^{\rm ADDA}$ outperforms TN$^{\rm OT}$ in constraining $\sigma_{8}$, which agrees with the obtained $R^2$ values in Table~\ref{tab:rsquare}.

We plot in Figure~\ref{fig:sim-tng} the features extracted by the source and the two adapted encoders from the HI SIMBA and HI IllustrisTNG maps respectively. For this two dimensional visualization, we opt for UMAP method \citep{mcinnes2018umap}. On the top row of Figure~\ref{fig:sim-tng}, red and blue dots denotes instances from in- and out-of-distribution datasets respectively. In the leftmost panel, we show representations of both SIMBA and IllustrisTNG maps extracted by the pre-trained source encoder $\bm{h}_{S}$. The shift in latent space between the source and target domains is noticeable. After adaptation of the target encoder $\bm{h}_{T}$, using either of the two methods (ADDA, optimal transport), the representations of SIMBA and IllustrisTNG obtained from $\bm{h}_{S}$ and $\bm{h}_{T}$ respectively are well aligned (see Figure~\ref{fig:sim-tng} middle and rightmost panels) as expected. This explains the reason why the regressor $\mathcal{R}_{S}$ pre-trained on the source domain is able to predict the underlying matter density of the test set from the out-of-distribution dataset. Each panel in the middle row of Figure~\ref{fig:sim-tng} is the same as the corresponding one in the top row but the color coding is based on the value of $\Omega_{\rm m}$. 
The difference between the leftmost panel and the other two is that the color gradient is more pronounced in the latter, indicating that the instances whose $\Omega_{\rm m}$ values are similar are well clustered in latent space. This trend also accounts for the fact that the maps are sensitive to the matter density. The color coding of all panels in the bottom row in Figure~\ref{fig:sim-tng} is related to the value of $\sigma_{8}$. Although not considerable, there is a slight difference between the leftmost panel (before adaptation) and the other two (after adaptation). The clustering of instances with similar underlying $\sigma_{8}$ values is the most pronounced in the middle panel corresponding to the ADDA method. This corroborates the results in Table~\ref{tab:rsquare} where the target network, adapted with ADDA, is able to recover $\sigma_{8}$ parameter from the out-of-distribution maps (i.e. IllustrisTNG maps) to an accuracy of  $R^2$ = 0.735.

\subsection{IllustrisTNG (IDs)$\rightarrow$SIMBA (OODs)}\label{subsec:tng-sim}
We now investigate the case where IllustrisTNG is considered as the source domain and SIMBA the target one (TNG$\rightarrow$SIMBA). After pre-training,
results show that the source network outperforms the one in SIMBA$\rightarrow$TNG case, as evidenced by its accuracy to recover the cosmological parameters with $R^{2}$ = 0.974, 0.884 for $\Omega_{\rm m}$ and $\sigma_{8}$ respectively (see also Table~\ref{tab:rsquare} second row). The constraint on $\sigma_{8}$ appears to be much tighter. The source network is also tested on the out-of-distribution examples and fails to constrain the underlying cosmology, as indicated by $R^{2}$ =  -2.109, -2.112 on $\Omega_{\rm m}$ and $\sigma_{8}$ respectively, as expected. We argue that the better performance of the source network to retrieve information from IllustrisTNG maps compared to those of SIMBA (during pre-training) is that the former exhibit much more details at high frequency modes (i.e. small scales) which also carry relevant information (see Figure~\ref{fig:images}). 
We find that, irrespective of the feature alignment method used (ADDA, OT), the adopted encoder $\bm{h}_{T}$ is able to constrain $\Omega_{\rm m}$
but fails to constrain $\sigma_{8}$ from SIMBA maps (see Table~\ref{tab:rsquare}). The $\Omega_{\rm m}$ predictions on out-of-distribution samples by trained $\bm{h}_{T}$ also appear to be slightly less precise than those obtained from the $\bm{h}_{S}$ on in-distribution instances, in contrast with the results that we have arrived at in the SIMBA$\rightarrow$TNG scenario. The overall somewhat underperformance of $\bm{h}_{T}$ in the TNG$\rightarrow$SIMBA case can be explained by the blurriness of the SIMBA maps on small scales which are relevant in order to extract the information about matter clustering that is directly related to $\sigma_{8}$. The challenge posed by extracting information from SIMBA as the target domain is the reason why the adaptation is run for more epochs. 

The large scatter related to $\sigma_{8}$ predictions in the bottom panels in Figure~\ref{fig:tng-sim-prediction} agrees well with the computed $R^2$ (0.205, -0.215) in Table~\ref{tab:rsquare}. Similar to the SIMBA$\rightarrow$TNG experiment, the density matter predictions (blue dots) correlates reasonably well with the ground truth. However, the larger dispersion at higher $\Omega_{\rm m}$ values ($>$ 0.3) corresponding to ADDA method could potentially explain the slightly lower performance of TN$^{\rm ADDA}$ ($R^2$ = 0.903) compared to that of TN$^{\rm OT}$ ($R^2$ = 0.924).

The trends, corresponding to the TNG$\rightarrow$SIMBA case, exhibited in Figure~\ref{fig:tng-sim}  are similar to those highlighted in Figure~\ref{fig:sim-tng}. There is an overlap, hence a good alignment, between representations of the source and target domains after adaptation of the target encoder $\bm{h}_{T}$ (see top row of Figure~\ref{fig:tng-sim}). The clustering of instances corresponding to similar $\Omega_{\rm m}$ values are also more apparent when representations of the out-of-distribution test set obtained from  trained $\bm{h}_{T}$ are shown together with those of the in-distribution test set (see middle row panels in Figure~\ref{fig:tng-sim}). The seemingly random clustering of points in latent space when using $\sigma_{8}$ value for the color coding (see Figure~\ref{fig:tng-sim} bottom panels) supports the findings in Table~\ref{tab:rsquare}. The target network, regardless of the adaptation method considered, fails to retrieve $\sigma_{8}$ from the out-of-distribution maps (i.e. SIMBA maps).

\subsection{Can we adapt the target encoder using fewer instances?}
In a real-world scenario, at an early stage of an HI experiment, we do not expect to collect thousands of maps for training a deep learning model. We therefore investigate the idea of adapting the target encoder using a relatively small number of instances from the target domain. It is worth reiterating that for this investigation, only the adaptation phase is run, and the target encoder is initialized and tested by using the pre-trained weights of the source encoder and the built regressor respectively. We adapt the target encoder using the two adaptation techniques in each setup, i.e. SIMBA$\rightarrow$TNG and TNG$\rightarrow$SIMBA, as before. The hyperparameters, including the number of epochs during the adaptation phase, in each setup for each technique are the same as those considered to arrive at the results presented in \S\ref{subsec:sim-tng} and \S\ref{subsec:tng-sim}. For this test, we select 100 instances from the target domain to train the target encoder. The cosmological predictions on out-of-distribution instances are presented in Table~\ref{tab:rsquare-small-samples}. It is clear that the constraints on the matter density are tighter than those of the amplitude of the density fluctuations in all cases. In each setup, the accuracy of the adapted network to extract the cosmological information from the target domain maps remains roughly the same using either ADDA or OT. On the one hand, the trained target network is still able to recover $\Omega_{\rm m}$ from the out-of-distribution maps in SIMBA$\rightarrow$TNG with a precision of $R^{2}$ $\sim$ 0.89. This is achieved despite the relatively small number of instances used to adapt it, about two orders of magnitude smaller than the training dataset used in \S\ref{subsec:sim-tng} (12,000 examples). The amplitude of the density fluctuations is recovered with an $R^{2}$ $\sim$ 0.61 using both methods. This provides further insights into the robustness of the techniques considered in this work. On the other hand, extracting the underlying matter density from the out-of-distribution maps in TNG$\rightarrow$SIMBA poses a challenge to the adapted network, consistent with the trend which has already been highlighted in \S\ref{subsec:tng-sim}. Regardless of the approach used, the trained target network achieves an $R^{2}\sim 0.69$ on $\Omega_{\rm m}$ overall. Inferring $\sigma_{8}$ is more challenging as indicated by $R^2$ = 0.11, 0.299 in ADDA and OT methods respectively. 

Interestingly when using OT method, although the target network fails to retrieve $\sigma_{8}$ in both experiments, the related $R^2$ values are higher than those obtained from using all instances (12,000) during adaptation phase (see Tables~\ref{tab:rsquare} and \ref{tab:rsquare-small-samples}). In the case of SIMBA$\rightarrow$TNG, this can be potentially explained by seemingly more pronounced clustering of points in Figure~\ref{fig:sim-tng-small} bottom row rightmost panel (compared to the corresponding panel in Figure~\ref{fig:sim-tng}). Similarly, in TNG$\rightarrow$SIMBA, the small clustering of $\sigma_{8}$ with high values (blue dots) in Figure~\ref{fig:tng-sim-small} bottom row rightmost panel (compared to that of Figure~\ref{fig:tng-sim} bottom row rightmost panel) might account for the difference in performance of TN$^{\rm OT}$. However, provided that TN$^{\rm OT}$ fails to constrain $\sigma_{8}$ in all cases, we can not conclude that TN$^{\rm OT}$ performs well when adapted with smaller dataset.


\section{Conclusions}\label{sec:conclusion}
We have shown in this work that extracting cosmological information from out-of-distribution HI maps is feasible by utilizing unsupervised domain adaptation techniques. We have considered SIMBA and IllustrisTNG HI maps, which are generated by two dissimilar distributions, owing to their own treatment of baryon physics. A source network which is composed of an encoder and a regressor is first pre-trained on the in-distribution dataset. The optimised weights of the source encoder serve as initial weights of a target encoder with identical architecture. We have opted for two different methods to align the features from the two domains in a lower dimension embedding space. An adversarial approach,  Adversarial Discriminative Domain Adaptation (ADDA), which resorts to a discriminator that differentiates the representations of the source domain from those of the target domain through the training. The second method is optimal transport which consists of minimizing the Wasserstein distance between the two representations. We note that the out-of-distribution dataset is assumed to be unlabeled. In other words, its labels (the cosmological parameters) are not accessed during the adaptation phase. The regressor is pre-trained along with the source encoder on the source domain and is directly used in combination with the adapted target encoder for inference on the test set from the target domain. The labels of the target domain are only utilized to evaluate the performance of the trained target encoder. We consider two scenarios where SIMBA and IllustrisTNG are the source and target domains respectively, i.e. SIMBA$\rightarrow$TNG, and vice versa, i.e. TNG$\rightarrow$SIMBA.

We find that in the SIMBA$\rightarrow$TNG case, the target encoder, which has been adapted using ADDA, achieves a performance comparable to that of the source encoder on the source domain. The prediction accuracies of the source network on $\Omega_{\rm m}$ and $\sigma_{8}$ are $R^{2}$ = 0.948 and 0.720 respectively on HI SIMBA dataset, whereas the trained target network achieves $R^{2}$ = 0.945, 0.735 ($\Omega_{\rm m}$, $\sigma_{8}$) on IllustrisTNG HI maps.
When using optimal transport for feature alignment, the target encoder fares equally well in terms of recovering $\Omega_{\rm m}$ from the HI IllustrisTNG maps, but fails to extract the amplitude of the density fluctuations $\sigma_{8}$. When cast into a two dimension subspace for visualization, the two representations of both source and target domains overlap well after feature alignment, as expected. Results show that, regardless of the adaptation method considered for the feature alignment, only the matter density can be retrieved from the out-of-distribution maps in the TNG$\rightarrow$SIMBA case. Extracting the amplitude of density fluctuations from the SIMBA maps proves to be quite challenging.

In the SIMBA$\rightarrow$TNG case, the weights of the pre-trained $\bm{h}_{S}$ are optimized to recover the cosmology from (relatively) blurry maps and are used to initialize the weights of $\bm{h}_{T}$ that are updated through feature alignment in order to predict the cosmology from more detailed maps of the target domain, i.e. IllustrisTNG. It is possible for the trained $\bm{h}_{T}$ to achieve a performance comparable to that of the source network, as indicated by the results in Table~\ref{tab:rsquare}. However, in the TNG$\rightarrow$SIMBA scenario, weights are first trained to retrieve information from relatively detailed source domain maps. They are then transferred to the target encoder weights that are adapted to infer cosmological parameters from target domain maps with less information in the high frequency modes. This makes the adaptation task more challenging, hence the asymmetry of the results from the two scenarios.

We have also explored the idea of utilizing relatively small dataset from the target domain, about 100 times smaller than the original target domain dataset in each setup, to adapt the target encoder. It has been found that, regardless of the size of the training samples used for the adaptation, the trained target network is still able to predict the matter density to a reasonable accuracy, $R^{2}\sim 0.89$, from IllustrisTNG HI maps in the SIMBA$\rightarrow$TNG setup. Extracting $\Omega_{\rm m}$ from SIMBA in TNG$\rightarrow$SIMBA, however, proves to be difficult.  

As the HI data that will be collected in the near future from various experiments will be unlabeled, the proof of concept described in this work plays a key role in terms of optimizing the information extraction from the maps. For future work, we will investigate the use of other more sophisticated technique such as Cycle-consistent adversarial domain adaptation \citep[Cycada;][]{hoffman2018cycada}, in order to further improve the robustness of the parameter inference on out-of-distribution samples. It is worth noting that our approach can be easily exploited for retrieving information from other large scale observables.\\

\begin{acknowledgments}
SA acknowledges financial support from the {\it South African Radio Astronomy Observatory} (SARAO). SH acknowledges support for Program number HST-HF2-51507 provided by NASA through a grant from the Space Telescope Science Institute, which is operated by the Association of Universities for Research in Astronomy, incorporated, under NASA contract NAS5-26555. The CAMELS project is supported by the Simons Foundation and NSF grant AST 2108078.  
\end{acknowledgments}

\bibliographystyle{mnras}
\bibliography{domain_adaptation}

\begin{thebibliography}{}
\makeatletter
\relax
\def\mn@urlcharsother{\let\do\@makeother \do\$\do\&\do\#\do\^\do\_\do\%\do\~}
\def\mn@doi{\begingroup\mn@urlcharsother \@ifnextchar [ {\mn@doi@}
  {\mn@doi@[]}}
\def\mn@doi@[#1]#2{\def\@tempa{#1}\ifx\@tempa\@empty \href
  {http://dx.doi.org/#2} {doi:#2}\else \href {http://dx.doi.org/#2} {#1}\fi
  \endgroup}
\def\mn@eprint#1#2{\mn@eprint@#1:#2::\@nil}
\def\mn@eprint@arXiv#1{\href {http://arxiv.org/abs/#1} {{\tt arXiv:#1}}}
\def\mn@eprint@dblp#1{\href {http://dblp.uni-trier.de/rec/bibtex/#1.xml}
  {dblp:#1}}
\def\mn@eprint@#1:#2:#3:#4\@nil{\def\@tempa {#1}\def\@tempb {#2}\def\@tempc
  {#3}\ifx \@tempc \@empty \let \@tempc \@tempb \let \@tempb \@tempa \fi \ifx
  \@tempb \@empty \def\@tempb {arXiv}\fi \@ifundefined
  {mn@eprint@\@tempb}{\@tempb:\@tempc}{\expandafter \expandafter \csname
  mn@eprint@\@tempb\endcsname \expandafter{\@tempc}}}

\bibitem[\protect\citeauthoryear{Alonso, Bull, Ferreira  \& Santos}{Alonso
  et~al.}{2015}]{alonso2015blind}
Alonso D.,  Bull P.,  Ferreira P.~G.,   Santos M.~G.,  2015, Monthly Notices of
  the Royal Astronomical Society, 447, 400

\bibitem[\protect\citeauthoryear{Andrianomena \& Hassan}{Andrianomena \&
  Hassan}{2023}]{andrianomena2023predictive}
Andrianomena S.,  Hassan S.,  2023, Journal of Cosmology and Astroparticle
  Physics, 2023, 051

\bibitem[\protect\citeauthoryear{Bandura et~al.,}{Bandura
  et~al.}{2014}]{bandura2014canadian}
Bandura K.,  et~al., 2014, in Ground-based and Airborne Telescopes V. pp
  738--757

\bibitem[\protect\citeauthoryear{Battye, Davies  \& Weller}{Battye
  et~al.}{2004}]{battye2004neutral}
Battye R.~A.,  Davies R.~D.,   Weller J.,  2004, Monthly Notices of the Royal
  Astronomical Society, 355, 1339

\bibitem[\protect\citeauthoryear{Battye, Browne, Dickinson, Heron, Maffei  \&
  Pourtsidou}{Battye et~al.}{2013}]{battye2013h}
Battye R.,  Browne I.,  Dickinson C.,  Heron G.,  Maffei B.,   Pourtsidou A.,
  2013, Monthly Notices of the Royal Astronomical Society, 434, 1239

\bibitem[\protect\citeauthoryear{Bharadwaj \& Pandey}{Bharadwaj \&
  Pandey}{2005}]{bharadwaj2005probing}
Bharadwaj S.,  Pandey S.~K.,  2005, Monthly Notices of the Royal Astronomical
  Society, 358, 968

\bibitem[\protect\citeauthoryear{Bigot-Sazy et~al.,}{Bigot-Sazy
  et~al.}{2015}]{bigot2015hi}
Bigot-Sazy M.-A.,  et~al., 2015, arXiv:1511.03006

\bibitem[\protect\citeauthoryear{Bonneel, Van De~Panne, Paris  \&
  Heidrich}{Bonneel et~al.}{2011}]{bonneel2011displacement}
Bonneel N.,  Van De~Panne M.,  Paris S.,   Heidrich W.,  2011, in Proceedings
  of the 2011 SIGGRAPH Asia conference. pp 1--12

\bibitem[\protect\citeauthoryear{Bousmalis, Trigeorgis, Silberman, Krishnan  \&
  Erhan}{Bousmalis et~al.}{2016}]{bousmalis2016domain}
Bousmalis K.,  Trigeorgis G.,  Silberman N.,  Krishnan D.,   Erhan D.,  2016,
  Advances in neural information processing systems, 29

\bibitem[\protect\citeauthoryear{Bull, Ferreira, Patel  \& Santos}{Bull
  et~al.}{2015}]{bull2015late}
Bull P.,  Ferreira P.~G.,  Patel P.,   Santos M.~G.,  2015, The Astrophysical
  Journal, 803, 21

\bibitem[\protect\citeauthoryear{Chang, Pen, Peterson  \& McDonald}{Chang
  et~al.}{2008}]{chang2008baryon}
Chang T.-C.,  Pen U.-L.,  Peterson J.~B.,   McDonald P.,  2008, Physical Review
  Letters, 100, 091303

\bibitem[\protect\citeauthoryear{{\'C}iprijanovi{\'c}, Kafkes, Jenkins, Downey,
  Perdue, Madireddy, Johnston  \& Nord}{{\'C}iprijanovi{\'c}
  et~al.}{2020}]{ciprijanovic2020domain}
{\'C}iprijanovi{\'c} A.,  Kafkes D.,  Jenkins S.,  Downey K.,  Perdue G.~N.,
  Madireddy S.,  Johnston T.,   Nord B.,  2020, arXiv:2011.03591

\bibitem[\protect\citeauthoryear{Ciprijanovic et~al.,}{Ciprijanovic
  et~al.}{2022}]{ciprijanovic2022deepadversaries}
Ciprijanovic A.,  et~al., 2022, Mach. Learn. Sci. Technol., 3, 35007

\bibitem[\protect\citeauthoryear{{\'C}iprijanovi{\'c}, Lewis, Pedro, Madireddy,
  Nord, Perdue  \& Wild}{{\'C}iprijanovi{\'c}
  et~al.}{2023}]{ciprijanovic2023deepastrouda}
{\'C}iprijanovi{\'c} A.,  Lewis A.,  Pedro K.,  Madireddy S.,  Nord B.,  Perdue
  G.~N.,   Wild S.~M.,  2023, Machine Learning: Science and Technology, 4,
  025013

\bibitem[\protect\citeauthoryear{Courty, Flamary, Tuia  \&
  Rakotomamonjy}{Courty et~al.}{2016}]{courty2016optimal}
Courty N.,  Flamary R.,  Tuia D.,   Rakotomamonjy A.,  2016, IEEE transactions
  on pattern analysis and machine intelligence, 39, 1853

\bibitem[\protect\citeauthoryear{Dav{\'e}, Angl{\'e}s-Alc{\'a}zar, Narayanan,
  Li, Rafieferantsoa  \& Appleby}{Dav{\'e} et~al.}{2019}]{dave2019simba}
Dav{\'e} R.,  Angl{\'e}s-Alc{\'a}zar D.,  Narayanan D.,  Li Q.,  Rafieferantsoa
  M.~H.,   Appleby S.,  2019, Monthly Notices of the Royal Astronomical
  Society, 486, 2827

\bibitem[\protect\citeauthoryear{DeBoer et~al.,}{DeBoer
  et~al.}{2017}]{deboer2017hydrogen}
DeBoer D.~R.,  et~al., 2017, Publications of the Astronomical Society of the
  Pacific, 129, 045001

\bibitem[\protect\citeauthoryear{Dell'Aiera, Vuillaume, Jacquemont  \&
  Benoit}{Dell'Aiera et~al.}{2023}]{dell2023deep}
Dell'Aiera M.,  Vuillaume T.,  Jacquemont M.,   Benoit A.,  2023, in
  Proceedings of the 20th International Conference on Content-based Multimedia
  Indexing. pp 133--139

\bibitem[\protect\citeauthoryear{Farahani, Voghoei, Rasheed  \&
  Arabnia}{Farahani et~al.}{2021}]{farahani2021brief}
Farahani A.,  Voghoei S.,  Rasheed K.,   Arabnia H.~R.,  2021, Advances in data
  science and information engineering: proceedings from ICDATA 2020 and IKE
  2020, pp 877--894

\bibitem[\protect\citeauthoryear{Faucher-Giguere, Lidz, Zaldarriaga  \&
  Hernquist}{Faucher-Giguere et~al.}{2009}]{faucher2009new}
Faucher-Giguere C.-A.,  Lidz A.,  Zaldarriaga M.,   Hernquist L.,  2009, The
  Astrophysical Journal, 703, 1416

\bibitem[\protect\citeauthoryear{Flamary et~al.,}{Flamary
  et~al.}{2021}]{flamary2021pot}
Flamary R.,  et~al., 2021, Journal of Machine Learning Research, 22, 1

\bibitem[\protect\citeauthoryear{Ganin \& Lempitsky}{Ganin \&
  Lempitsky}{2015}]{ganin2015unsupervised}
Ganin Y.,  Lempitsky V.,  2015, in International conference on machine
  learning. pp 1180--1189

\bibitem[\protect\citeauthoryear{Gilda, de Mathelin, Bellstedt  \&
  Richard}{Gilda et~al.}{2024}]{gilda2024unsupervised}
Gilda S.,  de Mathelin A.,  Bellstedt S.,   Richard G.,  2024, Unsupervised
  domain adaptation for constraining star formation histories

\bibitem[\protect\citeauthoryear{Gillet, Mesinger, Greig, Liu  \& Ucci}{Gillet
  et~al.}{2019}]{gillet2019deep}
Gillet N.,  Mesinger A.,  Greig B.,  Liu A.,   Ucci G.,  2019, Monthly Notices
  of the Royal Astronomical Society, 484, 282

\bibitem[\protect\citeauthoryear{Greig \& Mesinger}{Greig \&
  Mesinger}{2015}]{greig201521cmmc}
Greig B.,  Mesinger A.,  2015, Monthly Notices of the Royal Astronomical
  Society, 449, 4246

\bibitem[\protect\citeauthoryear{Haardt \& Madau}{Haardt \&
  Madau}{2012}]{haardt2012radiative}
Haardt F.,  Madau P.,  2012, The Astrophysical Journal, 746, 125

\bibitem[\protect\citeauthoryear{Hassan, Dav{\'e}, Finlator  \& Santos}{Hassan
  et~al.}{2017}]{hassan2017epoch}
Hassan S.,  Dav{\'e} R.,  Finlator K.,   Santos M.~G.,  2017, Monthly Notices
  of the Royal Astronomical Society, 468, 122

\bibitem[\protect\citeauthoryear{Hassan, Andrianomena  \& Doughty}{Hassan
  et~al.}{2020}]{hassan2020constraining}
Hassan S.,  Andrianomena S.,   Doughty C.,  2020, Monthly Notices of the Royal
  Astronomical Society, 494, 5761

\bibitem[\protect\citeauthoryear{Hassan et~al.,}{Hassan
  et~al.}{2022}]{hassan2022hiflow}
Hassan S.,  et~al., 2022, The Astrophysical Journal, 937, 83

\bibitem[\protect\citeauthoryear{Hoffman, Tzeng, Park, Zhu, Isola, Saenko,
  Efros  \& Darrell}{Hoffman et~al.}{2018}]{hoffman2018cycada}
Hoffman J.,  Tzeng E.,  Park T.,  Zhu J.-Y.,  Isola P.,  Saenko K.,  Efros A.,
   Darrell T.,  2018, in International conference on machine learning. pp
  1989--1998

\bibitem[\protect\citeauthoryear{Kantorovitch}{Kantorovitch}{1958}]{kantorovitch1958translocation}
Kantorovitch L.,  1958, Management science, 5, 1

\bibitem[\protect\citeauthoryear{Li et~al.,}{Li et~al.}{2023}]{li2023fast}
Li Y.,  et~al., 2023, The Astrophysical Journal, 954, 139

\bibitem[\protect\citeauthoryear{Liu \& Tegmark}{Liu \&
  Tegmark}{2011}]{liu2011method}
Liu A.,  Tegmark M.,  2011, Physical Review D—Particles, Fields, Gravitation,
  and Cosmology, 83, 103006

\bibitem[\protect\citeauthoryear{Liu, Tegmark, Bowman, Hewitt  \&
  Zaldarriaga}{Liu et~al.}{2009}]{liu2009improved}
Liu A.,  Tegmark M.,  Bowman J.,  Hewitt J.,   Zaldarriaga M.,  2009, Monthly
  Notices of the Royal Astronomical Society, 398, 401

\bibitem[\protect\citeauthoryear{Majumdar, Pritchard, Mondal, Watkinson,
  Bharadwaj  \& Mellema}{Majumdar et~al.}{2018}]{majumdar2018quantifying}
Majumdar S.,  Pritchard J.~R.,  Mondal R.,  Watkinson C.~A.,  Bharadwaj S.,
  Mellema G.,  2018, Monthly Notices of the Royal Astronomical Society, 476,
  4007

\bibitem[\protect\citeauthoryear{Makinen, Lancaster, Villaescusa-Navarro,
  Melchior, Ho, Perreault-Levasseur  \& Spergel}{Makinen
  et~al.}{2021}]{makinen2021deep21}
Makinen T.~L.,  Lancaster L.,  Villaescusa-Navarro F.,  Melchior P.,  Ho S.,
  Perreault-Levasseur L.,   Spergel D.~N.,  2021, Journal of Cosmology and
  Astroparticle Physics, 2021, 081

\bibitem[\protect\citeauthoryear{McInnes, Healy  \& Melville}{McInnes
  et~al.}{2018}]{mcinnes2018umap}
McInnes L.,  Healy J.,   Melville J.,  2018, arXiv preprint arXiv:1802.03426

\bibitem[\protect\citeauthoryear{Mellema et~al.,}{Mellema
  et~al.}{2013}]{mellema2013reionization}
Mellema G.,  et~al., 2013, Experimental Astronomy, 36, 235

\bibitem[\protect\citeauthoryear{Nelson et~al.,}{Nelson
  et~al.}{2019}]{nelson2019illustristng}
Nelson D.,  et~al., 2019, Computational Astrophysics and Cosmology, 6, 1

\bibitem[\protect\citeauthoryear{Papamakarios, Pavlakou  \&
  Murray}{Papamakarios et~al.}{2017}]{papamakarios2017masked}
Papamakarios G.,  Pavlakou T.,   Murray I.,  2017, Advances in neural
  information processing systems, 30

\bibitem[\protect\citeauthoryear{Pourtsidou, Bacon  \& Crittenden}{Pourtsidou
  et~al.}{2017}]{pourtsidou2017h}
Pourtsidou A.,  Bacon D.,   Crittenden R.,  2017, Monthly Notices of the Royal
  Astronomical Society, 470, 4251

\bibitem[\protect\citeauthoryear{Randrianjanahary, Karagiannis  \&
  Maartens}{Randrianjanahary et~al.}{2024}]{randrianjanahary2024cosmological}
Randrianjanahary L.~F.,  Karagiannis D.,   Maartens R.,  2024, Physics of the
  Dark Universe, 45, 101530

\bibitem[\protect\citeauthoryear{Rubner, Tomasi  \& Guibas}{Rubner
  et~al.}{2000}]{rubner2000earth}
Rubner Y.,  Tomasi C.,   Guibas L.~J.,  2000, International journal of computer
  vision, 40, 99

\bibitem[\protect\citeauthoryear{Santos et~al.,}{Santos
  et~al.}{2017}]{santos2017meerklass}
Santos M.~G.,  et~al., 2017, arXiv:1709.06099

\bibitem[\protect\citeauthoryear{Simonyan \& Zisserman}{Simonyan \&
  Zisserman}{2014}]{simonyan2014very}
Simonyan K.,  Zisserman A.,  2014, arXiv preprint arXiv:1409.1556

\bibitem[\protect\citeauthoryear{Smola, Gretton  \& Borgwardt}{Smola
  et~al.}{2006}]{smola2006maximum}
Smola A.~J.,  Gretton A.,   Borgwardt K.,  2006, in 13th international
  conference, ICONIP. pp~3--6

\bibitem[\protect\citeauthoryear{Sugiyama, Nakajima, Kashima, Buenau  \&
  Kawanabe}{Sugiyama et~al.}{2007}]{sugiyama2007direct}
Sugiyama M.,  Nakajima S.,  Kashima H.,  Buenau P.,   Kawanabe M.,  2007,
  Advances in neural information processing systems, 20

\bibitem[\protect\citeauthoryear{Sun \& Saenko}{Sun \&
  Saenko}{2016}]{sun2016deep}
Sun B.,  Saenko K.,  2016, in Computer Vision--ECCV 2016 Workshops: Amsterdam,
  The Netherlands, October 8-10 and 15-16, 2016, Proceedings, Part III 14. pp
  443--450

\bibitem[\protect\citeauthoryear{Tzeng, Hoffman, Saenko  \& Darrell}{Tzeng
  et~al.}{2017}]{tzeng2017adversarial}
Tzeng E.,  Hoffman J.,  Saenko K.,   Darrell T.,  2017, in Proceedings of the
  IEEE conference on computer vision and pattern recognition. pp 7167--7176

\bibitem[\protect\citeauthoryear{Villaescusa-Navarro
  et~al.,}{Villaescusa-Navarro et~al.}{2021}]{villaescusa2021camels}
Villaescusa-Navarro F.,  et~al., 2021, The Astrophysical Journal, 915, 71

\bibitem[\protect\citeauthoryear{Villaescusa-Navarro
  et~al.,}{Villaescusa-Navarro et~al.}{2022a}]{villaescusa2022camels1}
Villaescusa-Navarro F.,  et~al., 2022a, arxiv:2201.01300

\bibitem[\protect\citeauthoryear{Villaescusa-Navarro
  et~al.,}{Villaescusa-Navarro et~al.}{2022b}]{villaescusa2022camels}
Villaescusa-Navarro F.,  et~al., 2022b, The Astrophysical Journal Supplement
  Series, 259, 61

\bibitem[\protect\citeauthoryear{Wang et~al.,}{Wang
  et~al.}{2021}]{wang2021breakdown}
Wang Z.,  et~al., 2021, The Astrophysical Journal, 907, 4

\bibitem[\protect\citeauthoryear{Wold, Esbensen  \& Geladi}{Wold
  et~al.}{1987}]{wold1987principal}
Wold S.,  Esbensen K.,   Geladi P.,  1987, Chemometrics and intelligent
  laboratory systems, 2, 37

\makeatother
\end{thebibliography}

\appendix

\section{Architecture of the encoder}
To provide more information, we present in the Table~\ref{encoder_arch} the architecture of the encoder which is in both source and target networks.
\begin{table}
 \caption{The architecture of the encoder considered in our analyses. "BN + ReLU" denotes a batch normalization followed by a ReLU activation. The output shape is defined by the channel C, the height H and the width W. The batch size is the same throughout.}
 \begin{tabular}{lll}
  \hline
   & Layer & Output shape (C, H, W) \\
  \hline
  \hline
  1  & Input & (1, 256, 256)\\[2pt]
  2  & 3$\times$3 Convolutional Layer & (16, 256, 256)\\[2pt]
  3  & BN + ReLU & (16, 256, 256)\\[2pt]
  4  & 3$\times$3  Convolutional Layer & (16, 256, 256)\\[2pt]
  5 & BN + ReLU & (16, 256, 256)\\[2pt]
  6  & 2$\times$2  Max Pooling & (16, 128, 128)\\[2pt]
  7  & 3$\times$3  Convolutional Layer & (32, 128, 128)\\[2pt]
  8 & BN + ReLU & (32, 128, 128)\\[2pt]
  9  & 3$\times$3  Convolutional Layer & (32, 128, 128)\\[2pt]
  10  & BN + ReLU & (32, 128, 128)\\[2pt]
  11 & 2$\times$2 Max Pooling & (32, 64, 64)\\[2pt]
  12 & 3$\times$3  Convolutional Layer & (64, 64, 64)\\[2pt]
  13  & BN + ReLU & (64, 64, 64)\\[2pt]
  14 & 3$\times$3  Convolutional Layer & (64, 64, 64)\\[2pt]
  15  & BN + ReLU & (64, 64, 64)\\[2pt]
  16 & 3$\times$3  Convolutional Layer & (64, 64, 64)\\[2pt]
  17  & BN + ReLU & (64, 64, 64)\\[2pt]
  18 & 2$\times$2 Max Pooling & (64, 32, 32)\\[2pt]
  19 & 3$\times$3  Convolutional Layer & (128, 32, 32)\\[2pt]
  20  & BN + ReLU & (128, 32, 32)\\[2pt]
  21 & 3$\times$3  Convolutional Layer & (128, 32, 32)\\[2pt]
  22  & BN + ReLU & (128, 32, 32)\\[2pt]
  23 & 3$\times$3  Convolutional Layer & (128, 32, 32)\\[2pt]
  24  & BN + ReLU & (128, 32, 32)\\[2pt]
  25 & 2D adaptive average pooling & (128, 2, 2)\\[2pt]
  26 & Flattening & (512)\\[2pt]
 \hline
 \end{tabular}
 \label{encoder_arch}
\end{table}

\section{Feature alignment}
The architecture of the discriminator for the ADDA method is presented in the Table~\ref{discriminator_arch}. For the optimal transport method, the loss function simply uses the latent codes from both the source and target encoders to compute the Earth Mover's Distance.

\begin{table}
 \caption{The architecture of the discriminator used in the adversarial method, ADDA.}
 \begin{tabular}{lll}
  \hline
   & Layer & Output shape \\
  \hline
  \hline
  1  & Input & (512)\\[2pt]
  2  & Fully\ Connected\ Layer &  (512)\\[2pt]
  3  & ReLU & (512)\\[2pt]
  4  & Fully\ Connected\ Layer &  (512)\\[2pt]
  5 & ReLU & (512)\\[2pt]
  6  & Fully\ Connected\ Layer &  (512)\\[2pt]
  8 & ReLU & (512)\\[2pt]
  9  & Fully\ Connected\ Layer &  (2)\\[2pt]
  10  & LogSoftmax & (2)\\[2pt]
  \hline
 \end{tabular}
 \label{discriminator_arch}
\end{table}

\section{Features extracted: target encoder trained on small number of instances}
\begin{figure*}
\includegraphics[width=0.72\textwidth]{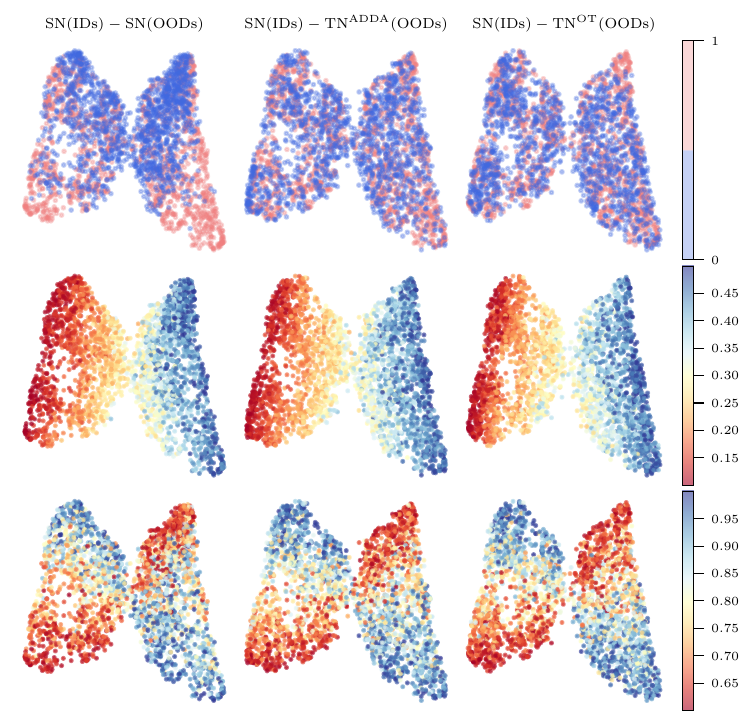}
\caption{\label{fig:sim-tng-small} Similar to what is shown in Figure~\ref{fig:sim-tng}, but the target encoder has been adapted with only 100 instances.}
\end{figure*}

\begin{figure*}
\includegraphics[width=0.72\textwidth]{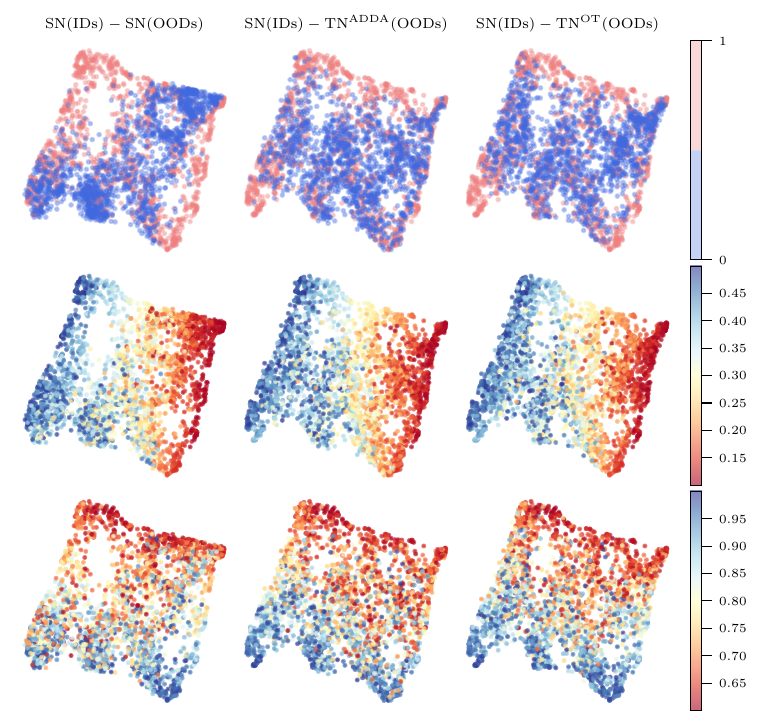}
\caption{\label{fig:tng-sim-small} Similar to what is shown in Figure~\ref{fig:tng-sim}, but the target encoder has been adapted with only 100 instances.}
\end{figure*}

For completeness, we are also plotting in Figures~\ref{fig:sim-tng-small} and \ref{fig:tng-sim-small} the distribution of the latent codes obtained from the target encoder when it is adapted using only 100 examples (similar to Figures~\ref{fig:sim-tng} and \ref{fig:tng-sim} respectively).




\end{document}